\documentclass[11pt,a4paper,english]{article}
\pdfoutput=1
\usepackage{graphicx}
\usepackage[T1]{fontenc}
\usepackage[utf8]{inputenc}
\usepackage{longtable}
\usepackage{amstext}
\usepackage{amssymb}

\makeatletter

\pdfpageheight\paperheight
\pdfpagewidth\paperwidth

\providecommand{\tabularnewline}{\\}

\usepackage{jcappub}
\usepackage{amsmath}
\usepackage{esint}
\pdfoutput=1

\renewcommand\[{\begin{equation}}
\renewcommand\]{\end{equation}}

\author[a]{Oton H. Marcori,}
\author[a]{Thiago S. Pereira}

\affiliation[a]{Departamento de Física, Universidade Estadual 
de Londrina, 86057-970, Londrina PR, Brazil}

\emailAdd{otonhm@hotmail.com}
\emailAdd{tspereira@uel.br}

\abstract{Two-point correlation functions are ubiquitous tools of modern 
cosmology, appearing in disparate topics ranging from cosmological inflation to 
late-time astrophysics. When the background spacetime is maximally symmetric, 
invariance arguments can be used to fix the functional dependence of 
this function as the invariant distance between any two points. 
In this paper we introduce a novel formalism which fixes
this functional dependence directly from the isometries
of the background metric, thus allowing one to quickly assess the overall
features of Gaussian correlators without resorting to the full machinery of perturbation 
theory. As an application we construct the CMB temperature correlation function in 
one inhomogeneous (namely, an off-center LTB model) and two spatially flat  
and anisotropic (Bianchi) universes, and derive their covariance matrices in the limit of 
almost Friedmannian symmetry. We show how the method can be extended to arbitrary 
$N$-point correlation functions and illustrate its use by constructing three-point correlation 
functions in some simple geometries.}


\makeatother

\usepackage{babel}
\begin{document}

\title{Two-point Correlation Functions in Inhomogeneous and Anisotropic
Cosmologies}
\maketitle

\section{Introduction}

A central assumption of the standard cosmological model is that the
universe we observe is a fair sample of an (hypothetical) ensemble
of universes. This hypothesis has far reaching consequences, but it
also brings along a whole statistical framework from which cosmological
observables are to be computed. It follows in particular that cosmological
parameters are not deduced directly from physical fields – which in
this framework are viewed as one realization of random variables –
but rather from their statistical moments, such as the one, two, and
higher $N$-point correlation functions. When using perturbation theory
to describe the clumpy universe, the one-point function is usually
defined to be zero, since one is actually interested in the fluctuations
of physical fields around their mean values. Thus, the first non-trivial
statistical moment is the two-point (or Gaussian) correlation function. 

Two-point functions are ubiquitous tools in modern physics. In field
theory they are disguised as Green's functions (or the propagator),
whereas in general relativity they could be simply a distance function
or a bitensor \cite{Allen:1985wd,Poisson:2011nh} – just to mention
a few examples. In cosmology, two-point \emph{correlation} functions
are a cornerstone of the standard $\Lambda$CDM model. Once it arises
as the quantization of a free field in the early inflationary universe
\cite{Linde:2005ht,Liddle:2000cg,Mukhanov:2005sc} (see ref. \cite{Uzan:2015pfm}
for an up-to-date review), it propagates to virtually all cosmological
and astrophysical computations one might be interested in – most popularly
in its Fourier (i.e., the power spectrum) version. The same reasoning
holds for higher-order correlation functions in connection with ``Beyond-$\Lambda$CDM''
approaches \cite{Abramo:2010gk,Bull:2015stt}. Therefore, knowledge
of the functional dependence of the two-point correlation function
(2pcf) is crucial, since it alone can tell a lot about the statistical
properties of cosmological observables, potentially allowing one to
disentangle cosmological signals from systematical effects in real
data. 

There are essentially two independent routes to find the functional
dependence of the 2pcf in cosmology. In the first, one uses heuristic
symmetry arguments (or its lack thereof) to fix this functional dependence.
This idea has been successfully applied in cosmology, mainly in connection
with CMB physics, in refs. \cite{Hajian:2003qq,Pullen:2007tu,Carroll:2008br,Pereira:2009kg,Abramo:2009fe,Froes:2015hva}.
However straightforward, the phenomenological quality of this approach
prevents one to link the resulting 2pcf to the statistics of a field
in a well-defined background geometry. Alternatively, one can deploy
the full machinery of perturbation theory in the desired spacetime.
After dealing with known issues of gauge invariance and mode decomposition,
the full set of Einstein equations can be solved and the statistics
of the 2pcf can be computed \cite{Pereira:2007yy,Clarkson:2009sc,pcm,Pitrou:2015iya,Pereira:2015pxa}.
This option is clearly more expensive, but is certain to lead to statistics
with known spacetime symmetries.

These considerations lead us to ask whether one can systematically
find the functional properties of correlation functions given the
spacetime symmetries, and without the need to resort to expensive
computations involving perturbation theory. In fact, when metric and
fluid perturbations are small, they can be seen as external fields
evolving over a fixed background, regardless of their dynamics. By
expanding such fields in an appropriate set of basis eigenfunctions,
one ensures that their statistical properties will inherit the symmetries
of the background metric. Thus, in a Friedmann-Lemaître-Robertson-Walker
(FLRW) spacetime, for example, the 2pcf of a random field can only
depend on the invariant distance between the two points, since this
is the only combination allowed by the symmetries of the FLRW metric.
Analogous ideas were explored in refs. \cite{Antoniadis:1996dj,Antoniadis:2011ib},
where the conformal invariance of the de Sitter spacetime has been
used to find the shape of two- and three-point correlation functions
in dark-energy dominated universes. 

In this work we systematically develop the idea of using the symmetries
of the background metric to fix the \emph{functional} form of the
2pcf. Starting from the definition of a two-point function in a general
manifold, we show in \S\ref{sec:Formalism} that the imposition of
isometric invariance on the 2pcf leads to a set of coupled first order
partial differential equations which can be solved by means of well
known techniques – but most easily through the method of characteristic
curves \cite{courant1966courant} – to fix the functional form of
the 2pcf. We illustrate the method in \S\ref{subsec:Homogeneous-and-isotropic},
where we show how it correctly recovers the 2pcf in spatially flat
FLRW spacetimes. We end this section by constructing in \S\ref{subsec:General-solution}
a formal solution to the aforementioned set of differential equations
which holds for any spacetime having at least one Killing vector.
In \S\ref{sec:Applications} we apply the formalism to obtain the
2pcf in two different classes of spacetimes. First, we consider the
case of an off-center inhomogeneous but spherically symmetric spacetime.
Then we show how the 2pcf will appear in a class of homogeneous but
spatially flat anisotropic geometries of the Bianchi family. In both
cases we derive the CMB temperature covariance matrix in the limit
of almost Friedmannian symmetry, and comment on their multipolar signatures.
In \S\ref{subsec:Non-Gaussian-correlations} we show how the method
can be easily generalized to include any $N$-point correlation function.
We conclude with some final remarks in \S\ref{sec:Final-remarks}.

\section{Formalism\label{sec:Formalism}}

We start with an informal description of what is meant by a two-point
correlation function in spacetime. For a rigorous and mathematically
complete description of two-point functions in Riemannian spaces,
see ref. \cite{kowalski1985two}. 

A \emph{two-point function} $f$ on a manifold ${\cal M}$ is simply
a real valued function of a pair of points $(p,q)\in{\cal M}\times{\cal M}$.
Known examples in physics are Green's functions, the geodesic distance
between two points or Synge's world function \cite{Poisson:2011nh}.
Here we shall be mainly interested in \emph{correlation} functions,
so we also demand $f$ to be symmetric
\[
f(p,q)=f(q,p),
\]
since correlation is clearly a pairwise concept. In most interesting
situations in cosmology one is dealing with the correlation of random
variables in spacetimes with some symmetries. Whenever these variables
can be viewed as external fields over a fixed background, their statistical
properties will inherit the symmetries of the underlying space. We
would thus like to define an invariant correlation function with respect
to these symmetries. Suppose that ${\cal M}$ possesses an isometry
represented by a one-parameter family of diffeomorphisms $\phi_{\tau}$
that maps any point $p\in{\cal M}$ to the point $\phi_{\tau}\left(p\right)\in{\cal M}$
such that $\phi_{0}(p)=p$. Clearly, $f$ will be invariant under
this symmetry if
\begin{equation}
f(p,q)=f\left(\phi_{\tau}(p),\phi_{\tau}(q)\right).\label{eq:2pcf-inv}
\end{equation}
In practice, though, one is always working in a specific coordinate
patch. Suppose that $\psi$ is a chart on an open interval of ${\cal M}$
and define 
\[
f\circ\psi^{-1}=f\left(\psi^{-1}(x_{1}^{\mu}),\psi^{-1}(x_{2}^{\mu})\right)\equiv\xi\left(x_{1}^{\mu},x_{2}^{\mu}\right).
\]
Therefore, the components of the curve $\phi_{\tau}$ in the coordinate
system defined by $\psi$ are
\[
\left.\left(\psi\circ\phi_{\tau}\right)^{\mu}\right|_{p}=x_{1}^{\mu}(\tau)\,,\qquad\text{and}\qquad\left.\left(\psi\circ\phi_{\tau}\right)^{\mu}\right|_{q}=x_{2}^{\mu}(\tau)\,.
\]
Locally, condition (\ref{eq:2pcf-inv}) then reads
\[
\xi(x_{1}^{\mu},x_{2}^{\mu})=\xi(x_{1}^{\mu}(\tau),x_{2}^{\mu}(\tau))
\]
which, for infinitesimal $\tau$, is equivalent to
\begin{equation}
\mathbf{K}\left(\xi\right)=0\,,\label{eq:2pcf-inv-2}
\end{equation}
where $\mathbf{K}=d/d\tau$ is a Killing vector, i.e., the vector
tangent to the curves generated by $\phi_{\tau}$. This condition
is nothing more than 
\[
\left.K^{\mu}\partial_{\mu}\xi\right|_{p}+\left.K^{\mu}\partial_{\mu}\xi\right|_{q}=0\,,\qquad K^{\mu}=\frac{dx^{\mu}}{d\tau}\,.
\]
Notice that in deriving this formula we are implicitly assuming that
both $p$ and $q$ are covered by the same coordinate system. Since
in general ${\cal M}$ can have several independent isometries, we
generalize the above result to the set of equations
\begin{equation}
\left.K_{\mathsf{a}}^{\mu}\partial_{\mu}\xi\right|_{p}+\left.K_{\mathsf{a}}^{\mu}\partial_{\mu}\xi\right|_{q}=0\,,\qquad\mathsf{a}\in\text{Isom}\left({\cal M}\right)\label{eq:2pcf-coord-inv}
\end{equation}
where Isom$\left({\cal M}\right)$ is the set of all isometries of
${\cal M}$. As we will see, this set of equations fully determine
the functional dependence of the 2pcf.

\subsection{Example: spatially flat FLRW universe\label{subsec:Homogeneous-and-isotropic}}

Equations (\ref{eq:2pcf-coord-inv}) form the core of our formalism.
They will lead to a set of coupled first order partial differential
equations which can be implicitly solved by means of the method of\emph{
characteristics curves} \cite{courant1966courant}. In order to illustrate
the method let us consider a two-point function in a spatially flat
FLRW universe; it could be, for example, the ensemble average of the
gravitational potential at two points on the same time slice. Since
FLRW universes are maximally symmetric expanding manifolds they possess
six independent Killing vectors: three of translation ($\mathbf{T}_{i}$)
and three of rotation ($\mathbf{R}_{i}$). In Cartesian coordinates
these vectors read 
\[
\mathbf{T}_{i}=\partial_{i}\,,\qquad\mathbf{R}_{i}=\epsilon_{ijk}x^{j}\partial^{k}\,.
\]
The two-point function depend on six variables: $\xi=\xi(x_{1},y_{1},\dots,z_{2})$.
In practice it is easier to work with $(\pm)$-coordinates defined
as 
\[
x_{\pm}=x_{2}\pm x_{1}\,,\qquad y_{\pm}=y_{2}\pm y_{1}\,,\qquad z_{\pm}=z_{2}\pm z_{1}\,,
\]
so that $\xi=\xi\left(x_{-},\dots,z_{+}\right)$. Let us start with
the vector $\mathbf{T}_{x}$. In Cartesian coordinates we have that
$\mathbf{T}_{x}=T_{x}^{\mu}\partial_{\mu}=\left(1,0,0\right)$, which
implies $T_{x}^{\mu}=\delta_{x}^{\mu}$. Thus, for this KV, equations
(\ref{eq:2pcf-coord-inv}) give:
\begin{equation}
2\frac{\partial\xi}{\partial x_{+}}=0\,.\label{eq:Tx-FLRW}
\end{equation}
Clearly, $\xi$ cannot depend on $x_{+}$. Since this conclusion will
not be straightforward in general, let us illustrate how it follows
from the method of characteristics. Let $\tau$ be the parameter along
the integral curves (i.e., the isometry) of $\mathbf{T}_{x}$. Thus,
by definition the tangent vector to this isometry is $\mathbf{T}_{x}=d/d\tau$,
and we have by virtue of eq. (\ref{eq:2pcf-inv-2}) that
\begin{equation}
\mathbf{T}_{x}(\xi)=\dot{x}_{-}\frac{\partial\xi}{\partial x_{-}}+\dot{x}_{+}\frac{\partial\xi}{\partial x_{+}}+\cdots+\dot{z}_{+}\frac{\partial\xi}{\partial z_{+}}=0\label{eq:dxidtau}
\end{equation}
where a dot means a (partial) derivative with respect to $\tau$.
Comparing (\ref{eq:Tx-FLRW}) and (\ref{eq:dxidtau}) we see that
all coordinates are constant along $\tau$ except for $x_{+}$. Therefore
$\xi$ cannot depend on it. A similar procedure using $\mathbf{T}_{y}$
and $\mathbf{T}_{z}$ tell us that $\xi$ cannot depend on either
$y_{+}$ or $z_{+}$, so that $\xi=\xi\left(x_{-},y_{-},z_{-}\right)$.
Consider next the vector $\mathbf{R}_{z}=d/d\rho$. Using $\mathbf{R}_{z}=\left(-y,x,0\right)$
on (\ref{eq:2pcf-coord-inv}) we find
\[
x_{-}\frac{\partial\xi}{\partial y_{-}}-y_{-}\frac{\partial\xi}{\partial x_{-}}=0\,.
\]
On the other hand, we also have that 
\[
\frac{d\xi}{d\rho}=\dot{x}_{-}\frac{\partial\xi}{\partial x_{-}}+\dot{y}_{-}\frac{\partial\xi}{\partial y_{-}}+\dot{z}_{-}\frac{\partial\xi}{\partial z_{-}}=0
\]
where a dot now stands for $\partial/\partial\rho$. By comparing
the last two equations we find 
\[
\dot{x}_{-}=-y_{-}\,,\qquad\dot{y}_{-}=x_{-}\,,\qquad\dot{z}_{-}=0\,.
\]
The first pair of equations can be easily decoupled, giving (after
an arbitrary choice of phase)
\begin{equation}
x_{-}=A\cos\rho\,,\qquad y_{-}=A\sin\rho\,,\label{eq:xy-flrw}
\end{equation}
where $A$ is not necessarily a constant, since it can depend on the
parameters of other isometries. The most general and $\rho$-independent
combination of $x_{-}$ and $y_{-}$ is\footnote{The variable $z_{-}$ is already $\rho$-independent, so it does not
enter into this combination.} $x_{-}^{2}+y_{-}^{2}=A^{2}$, so that $\xi=\xi\left(x_{-}^{2}+y_{-}^{2},z_{-}\right)$.
Moving on we now consider the vector $\mathbf{R}_{y}$. By the same
reasoning we find
\[
z_{-}\frac{\partial\xi}{\partial x_{-}}-x_{-}\frac{\partial\xi}{\partial z_{-}}=0\,.
\]
However we note that in virtue of (\ref{eq:xy-flrw}) $x_{-}$ and
$y_{-}$ are not independent anymore. We thus define $u_{-}^{2}\equiv x_{-}^{2}+y_{-}^{2}$
so that equation above becomes
\[
\frac{z_{-}x_{-}}{u_{-}}\frac{\partial\xi}{\partial u_{-}}-x_{-}\frac{\partial\xi}{\partial z_{-}}=0\,.
\]
If we now use $\mathbf{R}_{y}=d/dr$ and expand $d\xi/dr$ as a total
derivative we find by comparison that
\[
u_{-}\dot{u}_{-}=z_{-}x_{-}\,,\qquad\dot{z}_{-}=-x_{-}
\]
which is easily solved by
\[
u_{-}=B\sin r\,,\qquad z_{-}=B\cos r\,,
\]
with $B$ constant. Thus, a $\rho r$-independent combination of variables
is $x_{-}^{2}+y_{-}^{2}+z_{-}^{2}$, which incidentally tells us that
$A=B\sin r$. Introducing the notation $\mathbf{r}_{1,2}=\left(x_{1,2},y_{1,2},z_{1,2}\right)$
we finally conclude that
\begin{equation}
\xi\left(\left|\mathbf{r}_{2}-\mathbf{r}_{1}\right|\right)\equiv\xi_{FL}(r_{-})\label{eq:iso-2pcf}
\end{equation}
gives a solution to (\ref{eq:2pcf-coord-inv}) up to an overall power
of the argument. While we are on this topic, let us further remark
that for CMB large angle perturbations the above solution becomes
\begin{equation}
\xi_{FL}=\xi_{FL}\left(\Delta\eta\sqrt{2-2\cos\gamma}\right)=\sum_{\ell}\frac{2\ell+1}{4\pi}C_{\ell}P_{\ell}(\cos\gamma)\,,\label{eq:SW-2pcf}
\end{equation}
where $\gamma$ is the angle between $\mathbf{r}_{2}$ and $\mathbf{r}_{1}$
and $\Delta\eta$ is the distance to the last scattering surface.
Finally, note that there is no need to impose the condition $\mathbf{R}_{x}\left(\xi\right)=0$
since it is automatically ensured by the algebra of KVs\footnote{Physically, this steems from the fact that the constancy of two components
of the angular momentum vector implies the constancy of the third.}: $\mathbf{R}_{x}\left(\xi\right)=-\left[\mathbf{R}_{y},\mathbf{R}_{z}\right]\left(\xi\right)=0$. 

The method above also works for timelike isometries. For example,
in Minkowski spacetime one has, additionally to the vectors above,
three boost KVs and one time-translation KV. One can easily check
that the same procedure will give $\xi=\xi\left(\Delta s\right)$,
where $\Delta s=(\eta_{\mu\nu}x_{-}^{\mu}x_{-}^{\nu})^{1/2}$. In
the next section we will show that eq. (\ref{eq:2pcf-coord-inv})
can be formally solved for any spacetime having at least one isometry.
In proving this solution we will thus arrive at an independent formulation
of the problem which in some cases (most notably in Bianchi spacetimes)
is simpler than the above examples, and which also takes care of the
time dependence.
\begin{figure}
\begin{centering}
\includegraphics[scale=0.3]{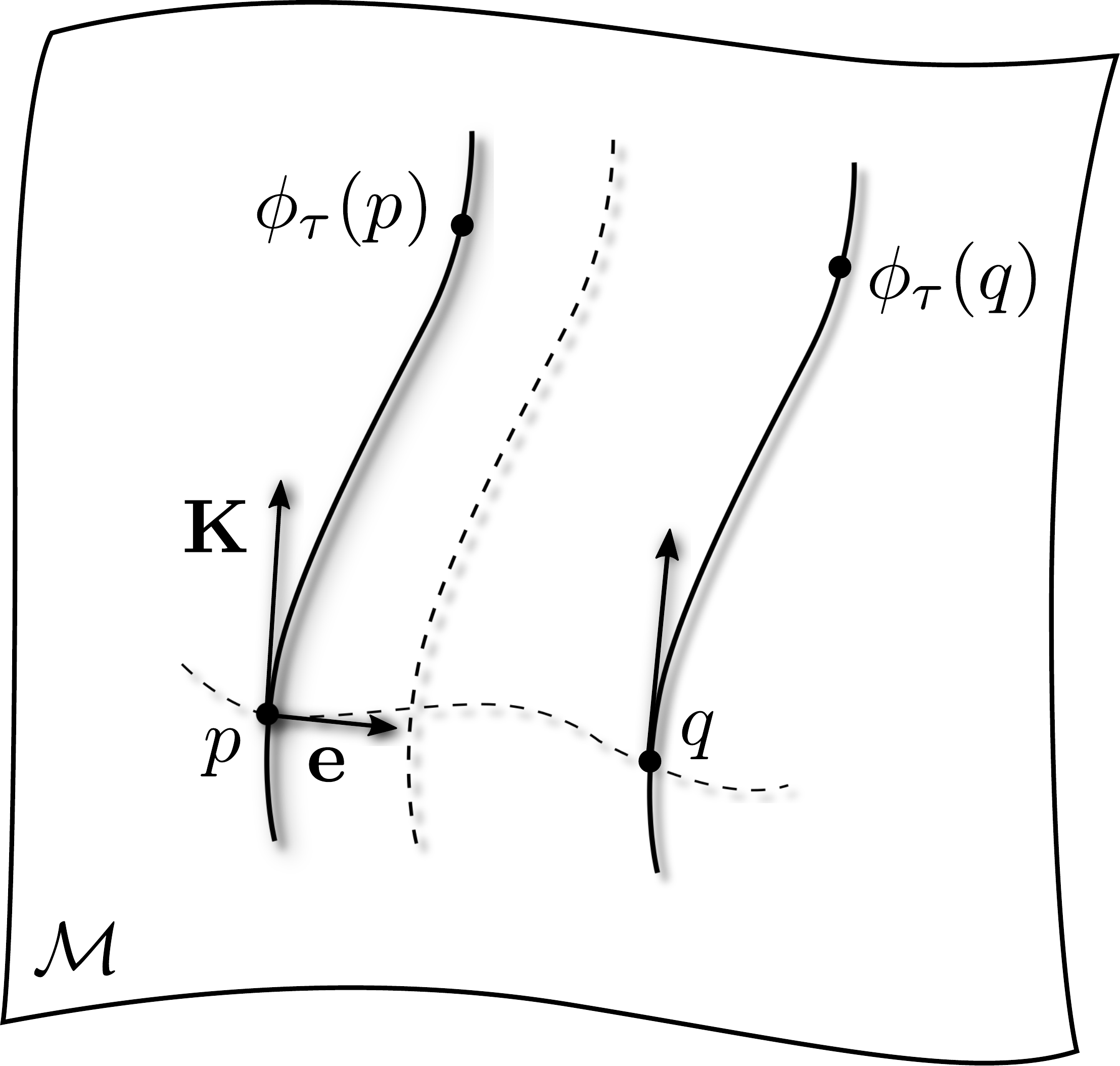}
\par\end{centering}
\caption{Schematic representation of two points on the manifold connected by
the vector $\mathbf{e}=d/ds$ and dragged by the Killing vector $\mathbf{K}=d/d\tau$.}

\label{fig:manifold}
\end{figure}

\subsection{General solution and time evolution\label{subsec:General-solution}}

As we have seen, the condition for a two-point correlation function
– or any two-point function – to be invariant with respect to an isometry
represented by the Killing vector $\mathbf{K}$, whose integral curves
are measured by a parameter $\tau$, is 
\begin{equation}
\frac{d\xi}{d\tau}=0\,.\label{eq:2pcf-inv-general}
\end{equation}
Suppose now that $\mathbf{e}$ is a vector field commuting with $\mathbf{K}$,
that is
\[
\pounds_{\mathbf{K}}e^{\mu}=\left[\mathbf{K},\mathbf{e}\right]^{\mu}=0\,.
\]
Then $\mathbf{e}$ is a vector connecting two close points on different
curves generated by $\mathbf{K}$. Moreover the quantity $u=g_{\mu\nu}e^{\mu}e^{\nu}$
is obviously constant along these isometries, since
\[
\pounds_{\mathbf{K}}u=\left(\pounds_{\mathbf{K}}g_{\mu\nu}\right)e^{\mu}e^{\nu}+2g_{\mu\nu}\left(\pounds_{\mathbf{K}}e^{\mu}\right)e^{\nu}=0
\]
where the first term is zero since $\mathbf{K}$ is a Killing vector.
This suggests that eq. (\ref{eq:2pcf-inv-general}) will be solved
for any function of $u$. That is
\[
\xi=\xi\left(\int_{s_{1}}^{s_{2}}\sqrt{g_{\mu\nu}e^{\mu}e^{\nu}}ds\right)\,,
\]
with $e^{\mu}=\partial x^{\mu}/\partial s$, is a solution of (\ref{eq:2pcf-inv-general}),
since $d\xi/d\tau=(d\xi/du)(du/d\tau)=0$. We have thus found a general
solution to eq. (\ref{eq:2pcf-inv-general}). Since in general we
will have more than one Killing vector, the 2pcf we will have more
then one argument, provided that we can find a set of independent
vectors $\left\{ \mathbf{e}_{i}\right\} $ commuting with the vectors
$\left\{ \mathbf{K}_{i}\right\} $. When this is the case
\[
\xi=\xi\left(\int_{r_{1}}^{r_{2}}\sqrt{g_{\mu\nu}e_{1}^{\mu}e_{1}^{\nu}}\,{\rm d}r,\int_{s_{1}}^{s_{2}}\sqrt{g_{\mu\nu}e_{2}^{\mu}e_{2}^{\nu}}\,{\rm d}s,\ldots\right)
\]
will be a solution to (\ref{eq:2pcf-coord-inv}). This solution is
particularly suited to the construction of 2pcf in homogeneous and
anisotropic Bianchi geometries where the basis $\{\mathbf{e}_{i}\}$
can always be constructed from the conditions $[\mathbf{K}_{i},\mathbf{e}_{j}]=0$.
In these cases the metric can be written as \cite{Hawgin-tetrad,Misner-terad}
\[
{\rm d}s^{2}=-{\rm d}\tau\otimes{\rm d}\tau+e^{2\alpha(\tau)}\left(e^{2\beta(\tau)}\right)_{ij}\mathbf{e}^{i}\otimes\mathbf{e}^{j}\,.
\]
Here the vectors $\{\mathbf{e}^{i}\}$ are the duals to $\{\mathbf{e}_{i}\}$,
$\left(e^{2\beta}\right)_{ij}$ is a symmetric and traceless $3\times3$
matrix whose eigenvalues are the directional scale factors, and $e^{\alpha}$
is the geometrically averaged scale factor. For these vectors we have
(no sum over $i$)
\[
g_{\mu\nu}e_{i}^{\mu}e_{j}^{\nu}=e^{2\alpha(\tau)}e^{2\beta_{ii}(\tau)}\delta_{ij}\,.
\]
This implies in particular that
\[
\int_{r_{1}}^{r_{2}}\sqrt{g_{\mu\nu}e_{1}^{\mu}e_{1}^{\nu}}{\rm d}r=e^{\alpha(\tau)}e^{\beta_{11}(\tau)}\left(r_{2}-r_{1}\right)
\]
with similar expressions for the other arguments. We have thus arrived
at a formal expression for the 2pcf which is valid in any Bianchi
spacetime 
\begin{equation}
\xi=\xi\left(e^{\alpha(\tau)}e^{\beta_{11}(\tau)}\left(r_{2}-r_{1}\right),e^{\alpha(\tau)}e^{\beta_{22}(\tau)}\left(s_{2}-s_{1}\right),e^{\alpha(\tau)}e^{\beta_{33}(\tau)}\left(t_{2}-t_{1}\right)\right)\,.\label{eq:general-2pcf-bianchi}
\end{equation}
To convert this function to one valid in an specific coordinate system
one have to find the parametric curves of the vectors $\mathbf{e}_{i}$
in the desired coordinates and invert these relations to obtain the
parameters as a function of the coordinates. Of course, the success
of this procedure depends on the coordinate system chosen. We will
illustrate this method with explicit examples in next section, where
we find $\xi$ for the geometries of Bianchi I and VII$_{0}$ universes.

\section{Applications\label{sec:Applications}}

We are now in position to put the above formalism to practical use.
We start in \S\ref{subsec:LTB} with the example of an inhomogeneous
universe with an off-center special point around which it is spherically
symmetric. This could be seen as an off-center LTB spacetime, though
in reality any spherically symmetric solution with a privileged point
will lead to the same answer. Then in \S\ref{subsec:Anisotropic-universes}
we consider two anisotropic spacetimes with spatially flat spatial
sections – namely, the models of Bianchi I and VII$_{0}$. We then
derive the Friedmannian limit of the 2pcf with first order corrections
in both cases, and connect the result with the temperature covariance
matrix of CMB fluctuations in \S\ref{subsec:CMB-covariance-matrix}.

\subsection{Universe with a special point\label{subsec:LTB}}

The 2pcf in an universe with a special point was studied from a phenomenological
standpoint in ref. \cite{Carroll:2008br}. More recently, the effect
of an off-center spherically symmetric void on the frequency and polarization
of CMB photons was investigated by the authors of ref. \cite{Cusin:2016kqx}.
Here we shall model an off-center spherically symmetric universe by
its Killing vectors. Let $\mathbf{w}=\left(a,b,c\right)$ represent
the spatial coordinates of this point with respect to our frame. Then
the only isometries are rotations about $\mathbf{w}$. These are represented
by the following KVs:
\[
\mathbf{R}_{i}=\epsilon_{ijk}\left(x^{j}-w^{j}\right)\partial^{k}\,.
\]
Let us start with rotations around the $z$-axis. Applying $\mathbf{R}_{z}=\left(-y+b,x-a,0\right)$
to eq. (\ref{eq:2pcf-coord-inv}) leads to
\[
\left(x_{+}-2a\right)\frac{\partial\xi}{\partial y_{+}}+x_{-}\frac{\partial\xi}{\partial y_{-}}-\left(y_{+}-2b\right)\frac{\partial\xi}{\partial x_{+}}-y_{-}\frac{\partial\xi}{\partial x_{-}}=0\,.
\]
Let $\rho$ be the parameter along the integral curves of $\mathbf{R}_{z}$,
such that $\mathbf{R}_{z}=d/d\rho$. Comparing the above with $d\xi/d\rho=0$
gives 
\[
\dot{x}_{+}=-y_{+}+2b\,,\quad\dot{x}_{-}=-y_{-}\,,\quad\dot{y}_{+}=x_{+}-2a\,,\quad\dot{y}_{-}=x_{-}\,,\quad\dot{z}_{\pm}=0\,.
\]
After decoupling and solving these equations we find that the combinations
$u_{-}^{2}\equiv x_{-}^{2}+y_{-}^{2}$ and $v_{+}\equiv\left(x_{+}-2a\right)^{2}+\left(y_{+}-2b\right)^{2}$
are constants with respect to $\rho$, so that $\xi=\xi\left(u_{-},v_{+},z_{-},z_{+}\right)$.
We next consider $\mathbf{R}_{y}=\left(z,0,-x\right)$ and change
variables from $\left(x_{-},x_{+}\right)$ to $\left(u_{-},v_{+}\right)$.
This gives
\[
\frac{z_{-}x_{-}}{u_{-}}\frac{\partial\xi}{\partial u_{-}}+\frac{\left(z_{+}-2c\right)\left(x_{+}-2a\right)}{v_{+}}\frac{\partial\xi}{\partial v_{+}}-\left(x_{+}-2a\right)\frac{\partial\xi}{\partial z_{+}}-x_{-}\frac{\partial\xi}{\partial z_{-}}=0\,.
\]
We now compare this to $d\xi/dr=0$, where $r$ is such that $\mathbf{R}_{y}=d/dr$.
This gives
\[
u_{-}\dot{u}_{-}=z_{-}x_{-}\,,\quad v_{+}\dot{v}_{+}=\left(z_{+}-2c\right)\left(x_{+}-2a\right),\quad\dot{z}_{+}=-x_{+}+2a\,,\quad\dot{z}_{-}=-x_{-}\,.
\]
Combining the last equation with the first and the third with the
second we find two constant combinations of variables: $u_{-}^{2}+z_{-}^{2}$
and $v_{+}^{2}+\left(z_{+}-2c\right)^{2}$. Thus, $\xi=\xi\left(u_{-}^{2}+z_{-}^{2},v_{+}^{2}+\left(z_{+}-2c\right)^{2}\right)$.
After a little algebra on the second argument, the final solution
can be written as
\[
\xi_{w}=\xi_{w}\left(\left|\mathbf{r}_{2}-\mathbf{r}_{1}\right|,\left|\mathbf{r}_{2}-\mathbf{w}\right|^{2}+\left|\mathbf{r}_{1}-\mathbf{w}\right|^{2}+2\left(\mathbf{r}_{2}-\mathbf{w}\right)\cdot\left(\mathbf{r}_{1}-\mathbf{w}\right)\right)\,.
\]
This result is compatible with the one found heuristically by the
authors of ref. \cite{Carroll:2008br}. Note however that the above
solution is more restrictive than theirs, since here we can obviously
write 
\begin{equation}
\xi_{w}=\xi_{w}\left(\left|\mathbf{r}_{2}-\mathbf{r}_{1}\right|,\left|\mathbf{r}_{2}+\mathbf{r}_{1}-2\mathbf{w}\right|\right)\,\label{eq:2pcf-single-point}
\end{equation}
whereas in \cite{Carroll:2008br} this is not possible\footnote{This difference does not affect the conclusions found by those authors,
since most of their results are actually extracted from an ansatz
of the power spectrum, and not from $\xi$.}. 

Before continuing we would like to make two comments about solution
(\ref{eq:2pcf-single-point}). First, the casual reader could be worried
that the above solution does not seem to recover (\ref{eq:iso-2pcf})
when $\mathbf{w}=0$. This happens because (\ref{eq:2pcf-single-point})
is only invariant under rotations, whereas (\ref{eq:iso-2pcf}) also
obeys translation symmetry. If the universe is homogeneous then $\mathbf{w}=0$
and we can further impose translation invariance through the condition
$\mathbf{T}_{\mathbf{r}}\left(\xi\right)=\nabla_{\mathbf{r}_{2}}\xi+\nabla_{\mathbf{r}_{1}}\xi=0$,
thus eliminating the dependence on $\left|\mathbf{r}_{2}+\mathbf{r}_{1}\right|$.
As a corollary of this result we find that the 2pcf in a inhomogeneous
but spherically symmetric universe about its origin – such as in Lemaître-Tolman-Bondi
(LTB) universes – is 
\begin{figure}
\begin{centering}
\includegraphics[scale=0.35]{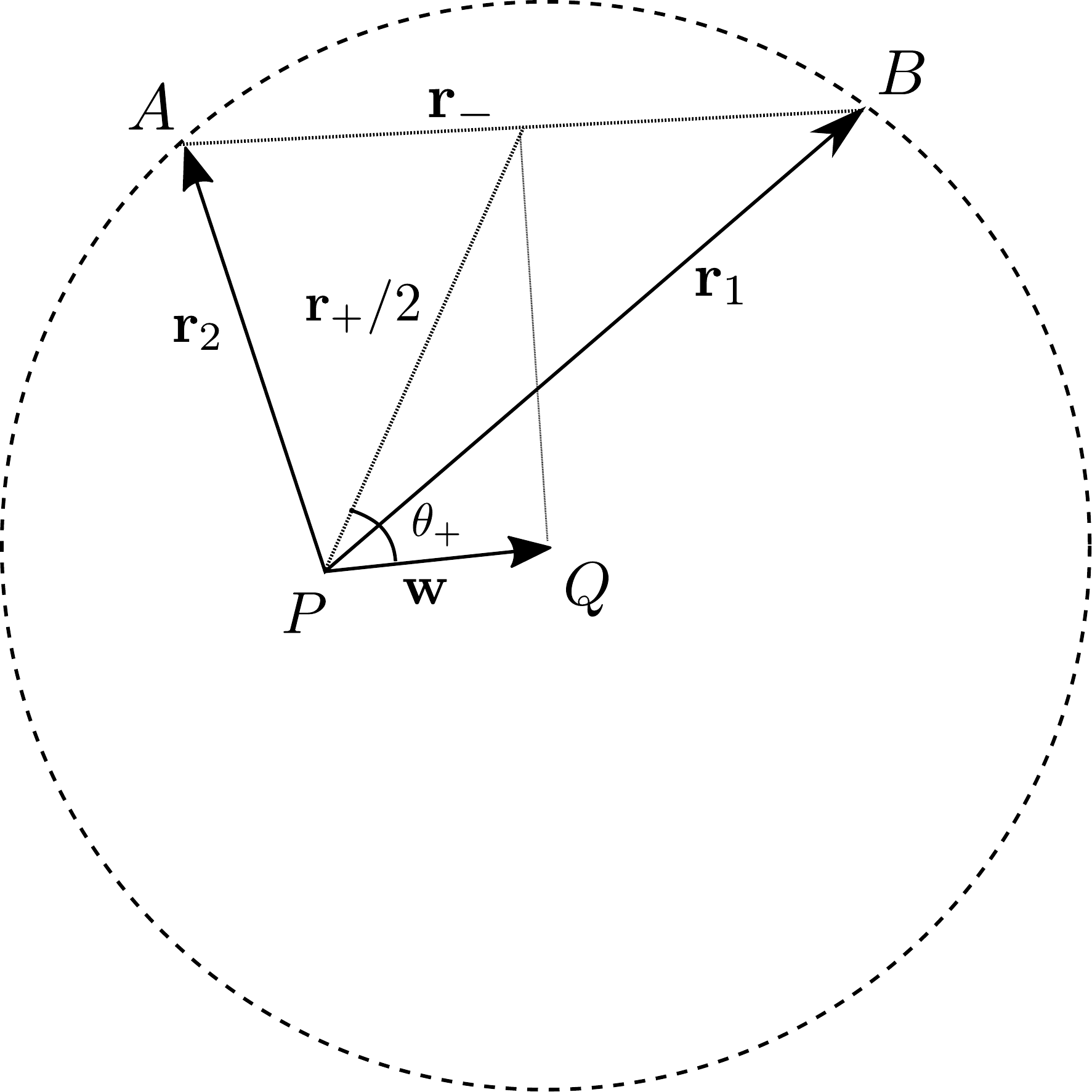}
\par\end{centering}
\caption{Schematic representation of a universe with a special point $Q$ at
a distance $\mathbf{w}$ from our position at $P$. In such universe
the correlation between two photons coming from positions $A$ and
$B$ can only depend on $\left|\mathbf{r}_{2}-\mathbf{r}_{1}\right|$
and $\left|\mathbf{r}_{2}+\mathbf{r}_{1}-2\mathbf{w}\right|$.}

\label{fig:off-center}
\end{figure}
\begin{equation}
\xi_{0}=\xi_{0}\left(\left|\mathbf{r}_{2}-\mathbf{r}_{1}\right|,\left|\mathbf{r}_{2}+\mathbf{r}_{1}\right|\right)\label{eq:ltb-2pcf}
\end{equation}
which is functionally equivalent to $\xi_{0}\left(r_{1},r_{2},\hat{\mathbf{r}}_{1}\cdot\hat{\mathbf{r}}_{2}\right)$,
since the only angle entering eq. (\ref{eq:ltb-2pcf}) is that between
$\mathbf{r}_{1}$ and $\mathbf{r}_{2}$. The second remark is that,
as stressed in \cite{Carroll:2008br}, the 2pcf (\ref{eq:2pcf-single-point})
possess a global shift symmetry of the form
\begin{equation}
\mathbf{r}_{1,2}\rightarrow\mathbf{r}_{1,2}+\mathbf{a}\,,\qquad\mathbf{w}\rightarrow\mathbf{w}+\mathbf{a}\label{eq:shift-symmetry}
\end{equation}
for any vector $\mathbf{a}$. This corresponds to the freedom in placing
a special point in an otherwise homogeneous universe, which is only
defined up to a global translation. We will come back to this issue
in \S\ref{subsec:Non-Gaussian-correlations}.

Equation (\ref{eq:2pcf-single-point}) concludes our task of finding
the 2pcf in an universe with a special point. The question of whether
we actually live close to an off-center spherically symmetric universe
can be tested by measuring off-diagonal terms in the covariance matrix
of CMB temperature fluctuations. Since we know that our universe is
very close to FLRW, we can test this hypothesis by deriving the FLRW
limit of eq. (\ref{eq:2pcf-single-point}), including leading order
corrections. To do that we first introduce the variables
\begin{equation}
\mathbf{r}_{\pm}=\mathbf{r}_{2}\pm\mathbf{r}_{1}\,.\label{eq:r_pluminus}
\end{equation}
Then we note that the desired limit of (\ref{eq:2pcf-single-point})
involves two independent expansions, namely, one in $\left|\mathbf{w}\right|$
and another one in powers of $r_{+}=\left|\mathbf{r}_{+}\right|$.
Let us start with the former. Assuming $\left|\mathbf{w}\right|\ll1$
we have
\begin{equation}
\left|\mathbf{r}_{+}-2\mathbf{w}\right|=r_{+}-2\hat{\mathbf{n}}_{+}\cdot\mathbf{w}+{\cal O}(\left|\mathbf{w}\right|^{2})\,.\label{eq:rplus-minus-2w}
\end{equation}
Thus
\begin{align}
\xi_{w} & \approx\xi_{w}(r_{-},r_{+}-2\hat{\mathbf{n}}_{+}\cdot\mathbf{w})\nonumber \\
 & =\xi_{0}\left(r_{-},r_{+}\right)-2\frac{\partial\xi_{0}\left(r_{-},r_{+}\right)}{\partial r_{+}}\hat{\mathbf{n}}_{+}\cdot\mathbf{w}+\cdots\,.\label{eq:LTB-perturb}
\end{align}
Next we assume that $\xi_{0}\left(r_{-},r_{+}\right)$ varies weakly
with $r_{+}$ and write
\begin{equation}
\xi_{0}\left(r_{-},r_{+}\right)=\xi_{0}\left(r_{-},0\right)+\frac{\partial\xi_{0}\left(r_{-},0\right)}{\partial r_{+}}r_{+}+\cdots\,.\label{eq:xi_0-expanded}
\end{equation}
It is important to note that we are not treating $r_{+}$ as a small
parameter. Indeed, this will hardly be the case, since for coincident
points on the CMB sphere we have $r_{+}=2\Delta\eta$, which is not
assumed as small. On the other hand, the assumption that $\xi_{0}$
varies weakly with $r_{+}$ implies that its translational invariance
is only slightly broken. That is
\[
\mathbf{T}_{\mathbf{r}}\left(\xi_{0}\left(r_{-},r_{+}\right)\right)=\nabla_{\mathbf{r}_{+}}\xi_{0}\left(r_{-},r_{+}\right)=\frac{\partial\xi_{0}\left(r_{-},0\right)}{\partial r_{+}}\hat{\mathbf{n}}_{+}\ll1\,.
\]
Since $\xi_{0}\left(r_{-},0\right)=\xi_{FL}(r_{-})$, we finally find
that
\begin{equation}
\xi_{w}=\xi_{FL}(r_{-})+\frac{\partial\xi_{0}\left(r_{-},0\right)}{\partial r_{+}}\left(r_{+}-2\hat{\mathbf{n}}_{+}\cdot\mathbf{w}\right)\,,\label{eq:2pcf-w-expanded}
\end{equation}
which is the desired result\footnote{Rigorously speaking (\ref{eq:2pcf-w-expanded}) should also include
second order terms since $\partial\xi_{0}/\partial r_{+}\left(\mathbf{n}_{+}\cdot\mathbf{w}\right)$
is formed from the product of two small quantities. However, second
order corrections from (\ref{eq:rplus-minus-2w}) will multiply $\partial\xi_{0}/\partial r_{+}$
in (\ref{eq:LTB-perturb}), producing a third order term. The only
remaining second order term is a correction to (\ref{eq:xi_0-expanded}),
but this does not induce any angular dependence on $\xi_{w}$.}.

In order to extract the amplitude of the leading corrections one still
needs the specific shape of the function $\xi_{0}\left(r_{-},r_{+}\right)$,
which at this point can only be fixed from first physical principles
\cite{Clarkson:2009sc,Masina:2010dc,Cusin:2016kqx}. Note however
that, as far as the angular dependence is concerned, there is no new
information in $\xi_{0}\left(r_{-},r_{+}\right)$ as compared to (\ref{eq:SW-2pcf}),
since its angular dependence also comes from the angle between $\mathbf{r}_{2}$
and $\mathbf{r}_{1}$. This means that the middle term in (\ref{eq:2pcf-w-expanded})
will not induce off-diagonal correlations in the CMB covariance matrix,
although it will surely alter the amplitude of the isotropic temperature
spectrum, i.e., the $C_{\ell}$s. On the other hand, the last term
will induce a dipole coming from the angle between $\hat{\mathbf{n}}_{+}$
and $\mathbf{w}$. We show in \S\ref{subsec:CMB-covariance-matrix}
how these multipolar coefficients can be directly linked to the temperature
covariance matrix.

\subsection{Anisotropic universes\label{subsec:Anisotropic-universes}}

In order to derive the 2pcf in Bianchi universes we start from the
general solution (\ref{eq:general-2pcf-bianchi}), which we have already
proven to solve (\ref{eq:2pcf-coord-inv}). One can check that the
same results follow instead from the direct application of eq. (\ref{eq:2pcf-coord-inv}),
up to the dependence on the directional scale factors. We will here
focus on two spatially flat anisotropic solutions and postpone a complete
analysis with other Bianchi metrics to a future work.

\subsubsection{Bianchi I}

We start with the simple Bianchi-I metric, which admits three translational
KVs:
\[
\mathbf{T}_{x}=\partial_{x}\,,\qquad\mathbf{T}_{y}=\partial_{y}\,,\qquad\mathbf{T}_{z}=\partial_{z}\,.
\]
The set of triad $\{\mathbf{e}_{i}\}$ vectors which are invariant
under the action of these isometries are \cite{stephani2009exact,Pontzen:2010eg}
\[
\mathbf{e}_{1}=\left(1,0,0\right)\,,\qquad\mathbf{e}_{2}=\left(0,1,0\right)\,,\qquad\mathbf{e}_{3}=\left(0,0,1\right)\,.
\]
Let us solve for the integral curves of the first vector. Putting
$e_{1}^{i}=dx^{i}/dr$ we find that $x=r$, $y=z=\text{constant}$.
Thus, we can invert the relation between the parameter and the coordinates
to find
\[
r_{2}-r_{1}=x_{2}-x_{1}=x_{-}\,.
\]
Solving for $\mathbf{e}_{2}$ and $\mathbf{e}_{3}$ leads to $s_{2}-s_{1}=y_{-}$
and $t_{2}-t_{1}=z_{-}$. Plugging these results back into (\ref{eq:general-2pcf-bianchi})
then gives 
\begin{equation}
\xi_{I}=\xi_{I}\left(e^{\alpha(\tau)}e^{\beta_{11}(\tau)}x_{-},e^{\alpha(\tau)}e^{\beta_{22}(\tau)}y_{-},e^{\alpha(\tau)}e^{\beta_{33}(\tau)}z_{-}\right)\,,\label{eq:2pcf-b1}
\end{equation}
which is the desired solution. 

Observational evidences tell us that our universe is very close to
isotropic \cite{Ade:2013nlj,Ade:2015hxq,Saadeh:2016sak}. We can thus
obtain the FLRW limit of $\xi_{I}$ by Taylor expanding this function
around $\beta_{ii}=0$. We find
\begin{equation}
\xi_{I}=\xi_{I}\left(x_{-},y_{-},z_{-}\right)+\left[\beta_{11}x_{-}\frac{\partial}{\partial x_{-}}+\beta_{22}y_{-}\frac{\partial}{\partial y_{-}}+\beta_{33}z_{-}\frac{\partial}{\partial z_{-}}\right]\xi_{I}\left(x_{-},y_{-},z_{-}\right)+\cdots
\end{equation}
where we have omitted the the functional dependence on $\alpha$ for
simplicity. In order to proceed, we note that the failure of the above
expression to be rotationally invariant is proportional to $\beta_{ii}$,
in the sense that full rotational isotropy should be exactly recovered
if $\beta_{ii}=0$. In fact, by applying, say, $\mathbf{R}_{z}$ to
the above expression we find
\[
\mathbf{R}_{z}\left(\xi_{I}\right)=\mathbf{R}_{z}\left(\xi_{I}\left(x_{-},y_{-},z_{-}\right)\right)+\beta_{11}\left(\cdots\right)+\beta_{33}\left(\cdots\right)+\beta_{33}\left(\cdots\right)
\]
where the ellipses contain terms like $x_{-}\partial_{y_{-}}\left(x_{-}\partial_{x_{-}}\xi\right)$
and so on, but which are not relevant for this discussion. The important
point is that the right hand side is linear in $\beta_{ii}$. Therefore,
for $\xi_{I}$ to be rotationally invariant at zero-order in $\beta_{ii}$,
it is necessary that $\mathbf{R}_{z}\left(\xi_{I}\left(x_{-},y_{-},z_{-}\right)\right)=0$
at this order. Repeating the analysis with $\mathbf{R}_{y}$ or $\mathbf{R}_{x}$
then leads to the isotropic condition 
\begin{equation}
\xi_{I}\left(x_{-},y_{-},z_{-}\right)=\xi_{FL}(r_{-})\,.\label{eq:iso-cond}
\end{equation}
Thus, the FLRW limit of (\ref{eq:2pcf-b1}) including first order
anisotropic corrections is 
\begin{equation}
\xi_{I}=\xi_{FL}+\frac{1}{r_{-}}\frac{\partial\xi_{FL}}{\partial r_{-}}\left[\beta_{11}x_{-}^{2}+\beta_{22}y_{-}^{2}+\beta_{33}z_{-}^{2}\right]\,.\label{eq:2pcf-b1-expanded}
\end{equation}
Clearly, this function will induce quadrupolar corrections in the
statistics of CMB, as is already known \cite{Barrow-CMBquadrupole,Gumrukcuoglu:2007bx,Pereira:2007yy,Pontzen:2010eg}.
Interestingly, though, it does not alter the isotropic spectrum, as
we will see in \S\ref{subsec:CMB-covariance-matrix}. 
\begin{figure}
\begin{centering}
\includegraphics[scale=0.9]{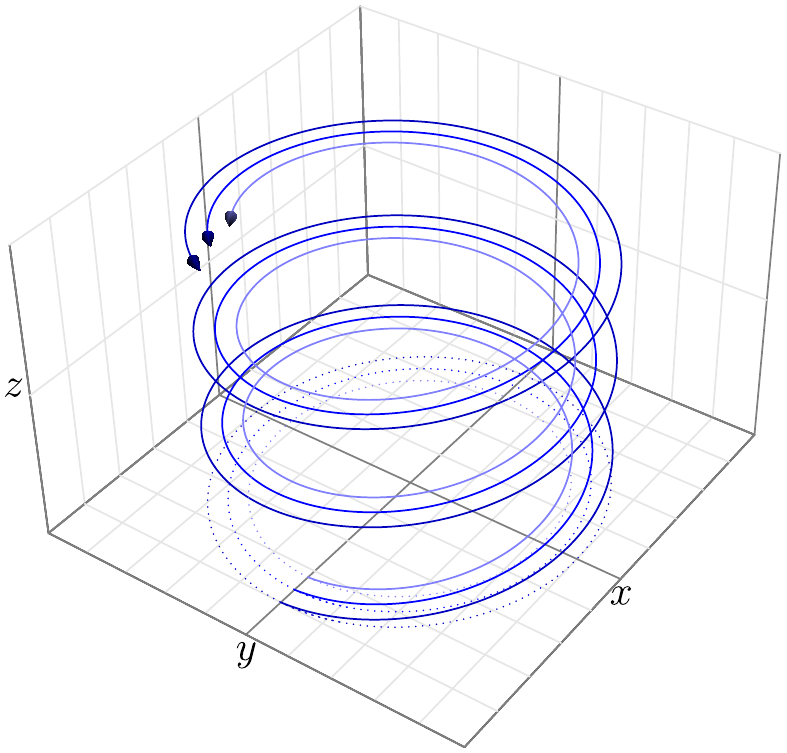}
\par\end{centering}
\caption{Integral curves of invariant vectors for Bianchi VII$_{0}$ spacetime.
These lead to quadrupole corrections in the covariance matrix, as
well as a change in the low-$\ell$ isotropic $C_{\ell}$s. See the
text for details.}

\label{fig:spirals}
\end{figure}

\subsubsection{Bianchi VII$_{0}$}

The spatial topology of Bianchi-VII$_{0}$ solution is $\mathbb{R}^{3}$,
so that it also has a flat FLRW limit when $\beta_{ii}=0$. The isometries
of this space can be seen as two orthogonal displacements in the $xy$-plane,
and a displacement in the $z$-axis followed by a rotation in the
$xy$-plane \cite{stephani2009exact,Pontzen:2010eg}. In Cartesian
coordinates the three KVs are\footnote{Note that we chose a different orientation of the axis as compared
to refs. \cite{stephani2009exact,Pontzen:2010eg}. }
\[
\mathbf{T}_{x}=\partial_{x}\,,\qquad\mathbf{T}_{y}=\partial_{y}\,,\qquad\mathbf{T}_{z}=\partial_{z}+x\partial_{y}-y\partial_{x}\,.
\]
The set of of invariant triad vectors are \cite{stephani2009exact,Pontzen:2010eg}
\[
\mathbf{e}_{1}=\left(\cos z,\sin z,0\right)\,,\quad\mathbf{e}_{2}=\left(-\sin z,\cos z,0\right)\,,\quad\mathbf{e}_{3}=\left(0,0,1\right).
\]
Let us consider the first vector with components $e_{1}^{i}=dx^{i}/dr$.
Its integral curves are
\[
x\left(r\right)=x_{0}+\left(\cos z_{0}\right)r\,,\qquad y(r)=y_{0}+\left(\sin z_{0}\right)r\,,\qquad z(r)=z_{0}\,.
\]
These relations can be easily inverted to give one of the curve segments
between the two points as a function of their coordinates: 
\[
r_{2}-r_{1}=\sqrt{x_{-}^{2}+y_{-}^{2}}\,.
\]
A similar computation involving $e_{2}^{i}=dx^{i}/ds$ and $e_{3}^{i}=dx^{i}/dt$
then gives the other two segments of the curve
\[
s_{2}-s_{1}=\sqrt{x_{-}^{2}+y_{-}^{2}}\,,\qquad t_{2}-t_{1}=z_{-}\,.
\]
This completes the task of finding the parameters of the integral
curves of the vectors $\{\mathbf{e}_{i}\}$ as a function of the coordinates.
Inserting the above expressions in (\ref{eq:general-2pcf-bianchi})
finally gives
\begin{equation}
\xi_{VII_{0}}=\xi_{VII_{0}}\left(e^{\alpha(\tau)}e^{\beta_{11}(\tau)}\sqrt{x_{-}^{2}+y_{-}^{2}},e^{\alpha(\tau)}e^{\beta_{22}(\tau)}\sqrt{x_{-}^{2}+y_{-}^{2}},e^{\alpha(\tau)}e^{\beta_{33}(\tau)}z_{-}\right)\,.\label{eq:2pcf-b7}
\end{equation}
The isotropic limit of $\xi_{VII_{0}}$ follows the same discussion
of the last section. The only difference is that, at zero-order in
$\beta_{ii}$, $\xi_{VII_{0}}$ is automatically invariant under $\mathbf{R}_{z}$,
as follows from its isometries. Thus we just require that $\mathbf{R}_{x}\left(\xi_{VII_{0}}\right)=0$
at zero order, which then gives
\[
\xi_{VII_{0}}\left(\sqrt{x_{-}^{2}+y_{-}^{2}},\sqrt{x_{-}^{2}+y_{-}^{2}},z_{-}\right)=\xi_{FL}\left(r_{-}\right).
\]
Thus
\begin{equation}
\xi_{VII_{0}}=\xi_{FL}(r_{-})+\frac{1}{r_{-}}\frac{\partial\xi_{FL}}{\partial r_{-}}\left[\left(\beta_{11}+\beta_{22}\right)\left(x_{-}^{2}+y_{-}^{2}\right)+\beta_{33}z_{-}^{2}\right]\label{eq:2pcf-b7-expanded}
\end{equation}
completes the desired expansion.

\subsection{CMB covariance matrix\label{subsec:CMB-covariance-matrix}}

Equations (\ref{eq:2pcf-single-point}), (\ref{eq:2pcf-b1}), and
(\ref{eq:2pcf-b7}), together with their FLRW expansions, are the
main results of last section. We emphasize that these results are
completely general and can be equally applied to large scale structures
as well as to CMB physics. Here we are interested in the latter, so
that we shall now derive the multipolar expansion of functions (\ref{eq:2pcf-w-expanded}),
(\ref{eq:2pcf-b1-expanded}), and (\ref{eq:2pcf-b7-expanded}) and
relate them to the corresponding CMB covariance matrix in the limit
of large angles, i.e., assuming only the Sachs-Wolfe effect.

We start by recalling that all the 2pcfs that we are considering have
the generic form
\begin{equation}
\xi(\mathbf{r}_{2},\mathbf{r}_{1})=\begin{cases}
\xi\left(\mathbf{r}_{-}\right) & \text{for Bianchi I and VII}_{0}\,,\\
\xi\left(\mathbf{r}_{+},\mathbf{w}\right) & \text{for a universe with a special point\,,}
\end{cases}\label{eq:2pcf-r1r2}
\end{equation}
where $\left(\mathbf{r}_{+},\mathbf{r}_{-}\right)$ were defined in
(\ref{eq:r_pluminus}) and \textbf{$\mathbf{w}$ }is the special point
introduced in \S\ref{subsec:LTB}. For simplicity, we have omitted
any extra dependence on $\left(r_{+},r_{-}\right)$, since these will
not lead to anisotropies. We will also omit the dependency of $\xi\left(\mathbf{r}_{+},\mathbf{w}\right)$
on $\mathbf{w}$, thus calling both 2pcfs above generically as $\xi(\mathbf{r}_{\pm})$.
This allows us to expand the $\xi(\mathbf{r}_{2},\mathbf{r}_{1})$
collectively as 
\[
\xi(\mathbf{r}_{\pm})=\sum_{\ell,m}\xi_{\ell m}(r_{\pm})Y_{\ell m}(\hat{\mathbf{n}}_{\pm})\,,\qquad\mathbf{r}_{\pm}=r_{\pm}\hat{\mathbf{n}}_{\pm}\,.
\]
Next we set
\[
\mathbf{r}_{\pm}=r_{\pm}\left(\sin\theta_{\pm}\cos\phi_{\pm},\sin\theta_{\pm}\cos\phi_{\pm},\cos\theta_{\pm}\right)\,,\qquad\mathbf{w}=w\hat{\mathbf{z}}\,,
\]
where $\theta_{+}=\arccos\left(\hat{\mathbf{n}}_{+}\cdot\hat{\mathbf{z}}\right)$,
and extract the coefficients $\xi_{\ell m}$. Notice that for $\mathbf{r}_{+}$
we have defined the $z$ axis along $\mathbf{w}$. The resulting expressions
are collected in Table \ref{Table1}, 
\begin{table}
\caption{Non-zero multipolar coefficients of anisotropic (Bianchi-I and Bianchi-VII$_{0}$)
and inhomogeneous (off-center LTB) 2pcfs considered in this work (respectively
eqs. (\ref{eq:2pcf-single-point}), (\ref{eq:2pcf-b1}), and (\ref{eq:2pcf-b7})).
The trace-free condition $\sum_{i}\beta_{ii}=0$ was used in all Bianchi
solutions. Note that the multipolar coefficients obey the reality
condition $\xi_{\ell m}^{*}=(-1)^{m}\xi_{\ell,-m}$. Primes in $\xi_{0}$
and $\xi_{FL}$ means $\partial/\partial r_{+}$ and $\partial/\partial r_{-}$,
respectively.}
\begin{longtable}{ccccc}
\hline
\hline 
Geometry & $\xi_{00}/\sqrt{4\pi}$ & $\xi_{10}$ & $\xi_{20}$ & $\xi_{22}$\tabularnewline
\hline 
\hline
\endhead
Off-center LTB & $\xi_{FL}+\xi_{0}'r_{+}$ & $-4w\sqrt{\frac{\pi}{3}}\xi'_{0}$ & 0 & 0\tabularnewline
Bianchi-I & $\xi_{FL}$ & 0 & $\beta_{33}\sqrt{\frac{4\pi}{5}}\,r_{-}\xi'_{FL}$ & $\left(\beta_{11}-\beta_{22}\right)\sqrt{\frac{2\pi}{15}}\,r_{-}\xi'_{FL}$\tabularnewline
Bianchi-VII$_{0}$ & $\xi_{FL}-\frac{1}{3}\beta_{33}\,r_{-}\xi'_{FL}$ & 0 & $\frac{4\beta_{33}}{3}\sqrt{\frac{4\pi}{5}}\,r_{-}\xi'_{FL}$ & 0\tabularnewline
\hline 
\end{longtable}\label{Table1}
\end{table}
 and can be directly related to the CMB temperature covariance matrix,
as we now show. 

At large scales the Sachs-Wolfe effect ($\Delta T/T=\Phi/3$) gives
the main contribution to the temperature fluctuations. In order to
compute the full effect of inhomogeneous or anisotropic geometries
in a real CMB map, gravitational evolution and re-ionization effects
should be taken into account. Clearly, such effects will not be provided
by our formalism, which is geometric in nature. On the other hand,
we can picture a scenario in which the asymmetries of the early universe
are washed out by inflation, but where quantum fluctuations preserve
such asymmetries on the statistics of the primordial gravitational
potential. This is the approach followed in, e.g., refs. \cite{Pereira:2007yy,Gumrukcuoglu:2007bx,Pullen:2007tu}.
In this scenario, primordial inhomogeneities and anisotropies are
contained in the statistics of CMB, and subsequent evolutionary effects
are assumed to be isotropic. Under this assumption the CMB covariance
matrix reads
\begin{equation}
\left\langle a_{\ell_{1}m_{1}}a_{\ell_{2}m_{2}}^{*}\right\rangle _{\pm}=\frac{1}{9}\int{\rm d}^{2}\mathbf{n}_{1}\int{\rm d}^{2}\mathbf{n}_{2}\left\langle \Phi(\mathbf{r}_{1})\Phi(\mathbf{r}_{2})\right\rangle _{\pm}Y_{\ell_{1}m_{1}}^{*}(\hat{\mathbf{n}}_{1})Y_{\ell_{2}m_{2}}(\hat{\mathbf{n}}_{2})\,.\label{eq:sw-effect}
\end{equation}
The two-point correlation function in this case is the ensemble average
of the gravitational potential: 
\[
\xi(\mathbf{r}_{\pm})=\left\langle \Phi(\mathbf{r}_{2})\Phi(\mathbf{r}_{1})\right\rangle _{\pm}\,.
\]
Once again, the $\pm$ labels correspond to the off-center LTB and
Bianchi models, respectively. Likewise, the $\pm$ notation in $\left\langle a_{\ell_{1}m_{1}}a_{\ell_{2}m_{2}}^{*}\right\rangle _{\pm}$
indicates that each covariance matrix corresponds to one of each correlation
function in (\ref{eq:2pcf-r1r2}). Usually, deviations from isotropy
and homogeneity are quantified directly in terms of the power spectrum
\[
_{\pm}P(\mathbf{k})=\int{\rm d}^{3}\mathbf{r}_{\pm}e^{-i\mathbf{k}\cdot\mathbf{r}_{\pm}}\xi\left(\mathbf{r}_{\pm}\right)\,.
\]
For example, by expanding $_{\pm}P(\mathbf{k})$ in harmonics one
can show that \cite{Pullen:2007tu,Abramo:2010gk} 

\begin{equation}
\left\langle a_{\ell_{1}m_{1}}a_{\ell_{2}m_{2}}\right\rangle _{\pm}=i^{\ell_{1}\pm\ell_{2}}\frac{2}{9\pi}\sum_{\ell,m}\int k^{2}{\rm d}k\,_{\pm}P_{\ell m}(k)j_{\ell_{1}}(k\Delta\eta)j_{\ell_{2}}(k\Delta\eta)(-1)^{m}{\cal G}_{-m,m_{1}m_{2}}^{\ell\ell_{1}\ell_{2}}\,,\label{eq:cov-matrix-powerspec}
\end{equation}
where
\[
{\cal G}_{m_{1}m_{2}m_{3}}^{\ell_{1}\ell_{2}\ell_{3}}=\int{\rm d}^{2}\hat{\mathbf{n}}\,Y_{\ell_{1}m_{1}}(\hat{\mathbf{n}})Y_{\ell_{2}m_{2}}(\hat{\mathbf{n}})Y_{\ell_{3}m_{3}}(\hat{\mathbf{n}})
\]
are the Gaunt coefficients (see Appendix \ref{subsec:gaunt}) and
$_{\pm}P_{\ell m}(k)$ are the multipolar coefficients of the power
spectrum. Then, from (\ref{eq:cov-matrix-powerspec}) and the coupling
properties of the Gaunt coefficients, a feature in the power spectrum
can be directly converted into a feature in the covariance matrix.
In fact it is easy to extract $_{\pm}P_{\ell m}(k)$ from the coefficients
in Table \ref{Table1} by means of the so-called Hankel transform
(see Appendix \ref{Appendix:hankel-Plm}):
\[
_{\pm}P_{\ell m}=4\pi i^{-\ell}\int_{0}^{\infty}r_{\pm}{\rm d}r_{\pm}\,j_{\ell}(kr_{\pm})\xi_{\ell m}(r_{\pm})\,.
\]
However, it is interesting to have an expression for the covariance
matrix directly in terms of $\xi_{\ell m}$. This can be obtained
by inserting the above expression into (\ref{eq:cov-matrix-powerspec}),
which gives
\begin{equation}
\left\langle a_{\ell_{1}m_{1}}a_{\ell_{2}m_{2}}\right\rangle _{\pm}=\frac{8}{9}\sum_{\ell_{3},m_{3}}\int_{0}^{2\Delta\eta}r_{\pm}^{2}{\rm d}r_{\pm}\,\xi_{\ell_{3}m_{3}}^{*}(r_{\pm})J_{\ell_{1}\ell_{2}\ell_{3}}^{(\pm)}(r_{\pm}){\cal G}_{m_{1}m_{2}m_{3}}^{\ell_{1}\ell_{2}\ell_{3}}\,,\label{eq:cov-matrix-real-space}
\end{equation}
where the coefficients $J_{\ell_{1}\ell_{2}\ell_{3}}^{(\pm)}$ are
implicitly defined in terms of the following integral 
\[
J_{\ell_{1}\ell_{2}\ell_{3}}^{(\pm)}(R,r_{1},r_{2})\equiv i^{\ell_{2}\pm\ell_{1}-\ell_{3}}\int_{0}^{\infty}k^{2}{\rm d}kj_{\ell_{1}}(kr_{1})j_{\ell_{2}}(kr_{2})j_{\ell_{3}}(kR)\,.
\]
This integral can be analytically solved and the result can be expressed
in terms of Wigner 6-J symbols \cite{Three-Bessel-int}. Parity symmetries
of the Wigner 6-J symbols then result in several properties of the
coefficients $J_{\ell_{1}\ell_{2}\ell_{3}}^{(\pm)}$ \cite{Three-Bessel-int}.
For our discussion, the most relevant property is that these coefficients
vanish whenever $R$ lies outside the range 
\[
\left|r_{1}-r_{2}\right|\leq R\leq r_{1}+r_{2}\,.
\]
In the context of CMB, $r_{1}=r_{2}=\Delta\eta$ and $r_{\pm}=\sqrt{2}\Delta\eta\sqrt{1\pm\hat{\mathbf{n}}_{2}\cdot\hat{\mathbf{n}}_{1}}$,
so that 
\[
0\leq r_{\pm}\leq2\Delta\eta\,.
\]
This explains why the domain of the integral in (\ref{eq:cov-matrix-real-space})
is limited. This result makes perfect sense since it is impossible
to consider points in the CMB sphere whose separation is larger than
$2\Delta\eta$. 

Returning to Table \ref{Table1} we see that each geometry leaves
its own fingerprint on the temperature spectrum. To the lowest order
in $\beta_{ii}$ in this formalism both Bianchi-I and VII$_{0}$ models
produce quadrupolar anisotropies whereas only the latter will alter
the isotropic temperature spectrum (i.e., the $C_{\ell}$s). Higher
multipoles come from higher order corrections in $\beta_{ii}$, but
we will not consider those here. Parity symmetry of these two models
prevent even-odd couplings of the harmonic coefficients \cite{Pereira:2015pga}.
For an off-center LTB universe, on the other hand, there will be dipolar
couplings as well as a change of the angular spectrum at low $\ell$s,
which depends on the derivatives of the function $\xi_{0}(r_{-},r_{+})$
evaluated at $r_{+}=0$. Although one can obtain these features by
inserting the coefficients $\xi_{\ell m}$ directly in (\ref{eq:cov-matrix-real-space}),
it is easier to relate them to the Bipolar Spherical Harmonics (BipoSH)
coefficients \cite{Hajian:2003qq,Hajian:2005jh} (see the Appendix
\ref{subsec:biposh})
\begin{equation}
^{(\pm)}{\cal A}_{\ell_{1}\ell_{2}}^{LM}=\frac{8}{9}\int r_{\pm}^{2}{\rm d}r_{\pm}\xi_{LM}(r_{\pm})J_{\ell_{1}\ell_{2}L}^{(\pm)}(r_{\pm}){\cal F}_{\ell_{1}\ell_{2}L}\,,\label{eq:biposh}
\end{equation}
where the set of coefficients ${\cal F}_{\ell_{1}\ell_{2}L}$ were
defined in (\ref{eq:Flll}), and relate these to the covariance matrix
using (\ref{eq:cov-from-biposh}). Thus, Bianchi-I and VII$_{0}$
geometries lead to a quadrupolar BipoSH ${\cal A}_{\ell_{1}\ell_{2}}^{2M}$,
whereas an off-center LTB produces a dipolar BipoSH ${\cal A}_{\ell_{1}\ell_{2}}^{1M}$.

As a final remark, we emphasize that expression (\ref{eq:cov-matrix-real-space})
should be seen as containing anisotropies and inhomogeneities only
from the initial conditions, after which we assume the universe to
be pure FLRW. In particular, contributions resulting from integrated
effects from the last scattering surface to us – like the effect induced
by the lensing potential in anisotropic \cite{Pereira:2015jya,Pitrou:2015iya}
and inhomogeneous \cite{Masina:2009wt,Masina:2010dc} universes –
cannot be extracted from this formalism in its present form. On the
other hand, the multipolar features resulting from the coefficients
in Table \ref{Table1} would still be preserved – perhaps in an integrated
version – as long as perturbations are functions of the background
coordinates. Indeed this is corroborated by the results of \cite{Pereira:2015jya,Pitrou:2015iya},
where quadrupolar corrections in the correlation of weak-lensing convergence
of large-scale structure in a Bianchi-I spacetime was found. Furthermore,
since (\ref{eq:2pcf-coord-inv}) was designed to work on scalar functions,
it cannot be directly applied to the cross-correlations between scalar,
vector and tensor perturbations, which are known to couple dynamically
through the evolution of the background shear \cite{Pitrou:2008gk,Gumrukcuoglu:2010yc}.
Nevertheless, since tensor fields are still seen as external fields
in a fixed background, tensor correlators should be expected to obey
a similar formalism as the one presented here (see also \cite{Allen:1985wd}).
We postpone a deeper investigation of these issues to a future work.

\section{Non-Gaussian correlations\label{subsec:Non-Gaussian-correlations}}

It is straightforward to extend this formalism to non-Gaussian correlation
functions. Let $\varphi$ be any $N$-point $\left(N>2\right)$ correlation
function. Repeating the arguments leading to condition (\ref{eq:2pcf-coord-inv})
then gives
\[
\sum_{j=1}^{N}\left.K_{\mathsf{a}}^{\mu}\partial_{\mu}\varphi\right|_{j}=0\,.
\]
The first non-trivial non-Gaussian statistical moment is the three-point
correlation function (3pcf). Let us consider this function in an FLRW
universe, where there are both translational and rotational symmetries.
Imposing invariance under the vector $T_{x}^{\mu}=\left(1,0,0\right)$
gives the following condition on $\varphi$:
\begin{equation}
\frac{\partial\varphi}{\partial x_{1}}+\frac{\partial\varphi}{\partial x_{2}}+\frac{\partial\varphi}{\partial x_{3}}=0\,.\label{eq:3pcf-x-translation}
\end{equation}
This is solved by any $\varphi$ with an arbitrary dependence on the
variable $X$ defined as 
\begin{equation}
X=lx_{1}+mx_{2}+nx_{3}\,,\qquad l+m+n=0\,,\label{eq:abc-constraint}
\end{equation}
with constants $\left(l,m,n\right)$. However, the constraint on these
constants allows us to write 
\[
X=l\left(x_{1}-x_{3}\right)+m\left(x_{2}-x_{3}\right),
\]
which shows that $\varphi$ can actually depend on the two ``base''
combinations $\left(x_{1}-x_{3}\right)$ and $\left(x_{2}-x_{3}\right)$.
Since we have no more constraints, these are the simplest combinations
of $x$-coordinates on which $\varphi$ can depend. Applying the same
reasoning for translations along $y$- and $z$-directions then gives
\[
\varphi_{\textrm{homog.}}\left(\mathbf{r}_{1},\mathbf{r}_{2},\mathbf{r}_{3}\right)=\varphi_{\textrm{homog.}}\left(\mathbf{r}_{1}-\mathbf{r}_{3},\mathbf{r}_{2}-\mathbf{r}_{3}\right)
\]
which is the most general homogeneous three-point function \cite{Pettinari:2014vja,Abramo:2010gk}.
To obtain an expression which is also invariant under rotations we
introduce $\mathbf{u}=\mathbf{r}_{1}-\mathbf{r}_{3}$ and $\mathbf{v}=\mathbf{r}_{2}-\mathbf{r}_{3}$
and simply note that the task of finding $\varphi\left(\mathbf{u},\mathbf{v}\right)$
invariant under rotations has already been solved in \S\ref{subsec:LTB}.
The solution is simply a function depending on the modulus of $\mathbf{u}\pm\mathbf{v}$
(see eqs. (\ref{eq:ltb-2pcf})). In terms of the original variables
this becomes\footnote{Note that since $\varphi\left(\left|\mathbf{u}-\mathbf{v}\right|,\left|\mathbf{u}+\mathbf{v}\right|\right)$
is equivalent to $\varphi\left(u,v,\mathbf{u}\cdot\mathbf{v}\right)$,
eq. (\ref{eq:3pcf-flrw}) is also equivalent to $\varphi\left(\left|\mathbf{r}_{1}-\mathbf{r}_{3}\right|,\left|\mathbf{r}_{2}-\mathbf{r}_{3}\right|,\left(\mathbf{r}_{1}-\mathbf{r}_{3}\right)\cdot\left(\mathbf{r}_{2}-\mathbf{r}_{3}\right)\right)$,
which appears to be more common in the literature \cite{Pettinari:2014vja}.}
\begin{equation}
\varphi_{FL}=\varphi_{FL}\left(\left|\mathbf{r}_{1}-\mathbf{r}_{2}\right|,\left|\mathbf{r}_{1}+\mathbf{r}_{2}-2\mathbf{r}_{3}\right|\right)\,.\label{eq:3pcf-flrw}
\end{equation}
Since the reasoning we used to arrive at this result might not be
entirely obvious, we note that rotations around the $z$-axis of the
vectors $\mathbf{r}_{1}$, $\mathbf{r}_{2}$ and $\mathbf{r}_{3}$
are equal to rotations around the $z$-axis of $\mathbf{u}$ and $\mathbf{v}$:
\begin{align*}
\mathbf{R}_{z} & =\left(x_{1}\frac{\partial}{\partial y_{1}}-y_{1}\frac{\partial}{\partial x_{1}}\right)+\left(x_{2}\frac{\partial}{\partial y_{2}}-y_{2}\frac{\partial}{\partial x_{2}}\right)+\left(x_{3}\frac{\partial}{\partial y_{3}}-y_{3}\frac{\partial}{\partial x_{3}}\right)\\
 & =x_{1}\frac{\partial}{\partial u_{y}}-y_{1}\frac{\partial}{\partial u_{x}}+x_{2}\frac{\partial}{\partial v_{y}}-y_{2}\frac{\partial}{\partial v_{x}}+x_{3}\left(-\frac{\partial}{\partial u_{y}}-\frac{\partial}{\partial v_{y}}\right)-y_{3}\left(-\frac{\partial}{\partial u_{x}}-\frac{\partial}{\partial v_{x}}\right)\\
 & =\left(u_{x}\frac{\partial}{\partial u_{y}}-u_{y}\frac{\partial}{\partial u_{x}}\right)+\left(v_{x}\frac{\partial}{\partial v_{y}}-v_{y}\frac{\partial}{\partial v_{x}}\right)\,,
\end{align*}
where we have introduced a (hopefully obvious) new notation for the
components of $\mathbf{u}$ and $\mathbf{v}$. An equivalent result
holds for $\mathbf{R}_{y}$ and $\mathbf{R}_{x}$, as one can easily
check. Then, by repeating the analysis of \S\ref{subsec:LTB} we
find $\varphi=\varphi\left(\left|\mathbf{u}-\mathbf{v}\right|,\left|\mathbf{u}+\mathbf{v}\right|\right)$
which gives (\ref{eq:3pcf-flrw}) upon replacing $\mathbf{u}$ and
$\mathbf{v}$ by their definitions. 

There is one interesting remark we would like to make about eq. (\ref{eq:3pcf-flrw}).
Notice that if we make the identification $\mathbf{r}_{3}=\mathbf{w}$
the 3pcf will have exactly the same functional dependence as the 2pcf
in eq. (\ref{eq:2pcf-single-point}) – namely, a Gaussian correlation
in an universe with a special point. This suggests that the bispectrum
(the Fourier transform of the 3pcf) in a FLRW universe could mimic
the power spectrum in a off-center LTB universe. Interestingly, it
has been argued that a (statistically homogeneous and isotropic) bispectrum
in the strong squeezed limit will induce statistical anisotropies
in the power spectrum \cite{Schmidt:2012ky}. In Fourier space the
power spectrum and bispectrum have the form (assuming statistical
homogeneity and isotropy)
\begin{align}
\xi(\mathbf{k}_{1},\mathbf{k}_{2}) & =P(k_{1})\delta\left(\mathbf{k}_{1}+\mathbf{k}_{2}\right)\,,\label{eq:pspec-fourier}\\
\varphi(\mathbf{k}_{1},\mathbf{k}_{2},\mathbf{k}_{3}) & =B(k_{1},k_{2},\mathbf{k}_{1}\cdot\mathbf{k}_{2})\delta\left(\mathbf{k}_{1}+\mathbf{k}_{2}+\mathbf{k}_{3}\right)\,.\label{eq:bispec-fourier}
\end{align}
In the squeezed limit $\mathbf{k}_{1}\approx-\mathbf{k}_{2}$ the
wave vector $\mathbf{k}_{3}\approx0$ corresponds to a long wavelength
perturbation which is equivalent to a spatial gradient. This gradient
modulates the lower order statistics leading to an effective power
spectrum which is now anisotropic: $P(k_{1})\rightarrow P_{\textrm{eff}}(\mathbf{k}_{1})$.
We add to these the fact that the delta $\delta\left(\mathbf{k}_{1}+\mathbf{k}_{2}+\mathbf{k}_{3}\right)$
in the bispectrum breaks the statistical independence previously existing
between $\mathbf{k}_{1}$ and $\mathbf{k}_{2}$ in (\ref{eq:pspec-fourier}).
Thus, in the presence of a bispectrum $\varphi(\mathbf{k}_{1},\mathbf{k}_{2},\mathbf{k}_{3})$,
$\xi(\mathbf{k}_{1},\mathbf{k}_{2})$ is no longer translational invariant\footnote{Note that this holds for any $\mathbf{k}_{1}$, $\mathbf{k}_{2}$
and $\mathbf{k}_{3}$, regardless of the squeezed limit.}. In real space, the similarity between (\ref{eq:3pcf-flrw}) and
(\ref{eq:2pcf-single-point}) is just reflecting the fact that the
third point in the 3pcf could itself be seen as a ``special'' point.
Analogously, a special point of an LTB universe will itself correlate
with any two points previously correlated.

As one last application let us consider the 3pcf in a LTB universe.
Rotational invariance around $R_{z}^{\mu}=\left(-y,x,0\right)$ gives
\[
\left(x_{1}\frac{\partial\varphi}{\partial y_{1}}+x_{2}\frac{\partial\varphi}{\partial y_{2}}+x_{3}\frac{\partial\varphi}{\partial y_{3}}\right)-\left(y_{1}\frac{\partial\varphi}{\partial x_{1}}+y_{2}\frac{\partial\varphi}{\partial x_{2}}+y_{3}\frac{\partial\varphi}{\partial x_{3}}\right)=0\,.
\]
We could try solving this equation with the introduction of two new
variables $X=lx_{1}+mx_{2}+nx_{3}$ and $Y=ly_{1}+my_{2}+ny_{3}$.
This would give
\[
X\frac{\partial\varphi}{\partial Y}-Y\frac{\partial\varphi}{\partial X}=0\,.
\]
The use of characteristics would then tell us that $\dot{X}=-Y$ and
$\dot{Y}=X$, which implies that $\varphi$ is a function of the constant
combination $X^{2}+Y^{2}$. This solution however is not the most
general one. To see that, note that in the absence of translational
invariance the constraint in (\ref{eq:abc-constraint}) no longer
holds. In this case we have
\[
l+m+n=2p
\]
for some constant $p$. We can thus rewrite the variable $X$ as
\[
X=l\left(x_{1}-x_{3}\right)+m\left(x_{2}-x_{3}\right)+p\left(x_{1}+x_{3}\right)+p\left(x_{2}+x_{3}\right)-p\left(x_{1}+x_{2}\right)
\]
with an analogous expression for $Y$. This tell us that there are
actually five ``base'' combinations on which $\varphi$ will depend,
i.e., $\varphi=\varphi\left(x_{1}-x_{3},x_{2}-x_{3},\dots,x_{1}+x_{2},\dots\right)$.
Repeating the analysis for $\mathbf{R}_{y}$ and $\mathbf{R}_{x}$,
which we hope by now has become clear, we find
\begin{equation}
\varphi_{0}=\varphi_{0}\left(\left|\mathbf{r}_{1}-\mathbf{r}_{3}\right|,\left|\mathbf{r}_{2}-\mathbf{r}_{3}\right|,\left|\mathbf{r}_{1}+\mathbf{r}_{3}\right|,\left|\mathbf{r}_{2}+\mathbf{r}_{3}\right|,\left|\mathbf{r}_{1}+\mathbf{r}_{2}\right|\right)\,.\label{eq:3pcf-ltb}
\end{equation}
Note in particular that the combination $\left|\mathbf{r}_{1}+\mathbf{r}_{2}\right|$
cannot be neglected, as one could have expected from a naive comparison
with (\ref{eq:ltb-2pcf}). The reason is that while $\mathbf{r}_{1}-\mathbf{r}_{2}$
is linearly dependent on $\mathbf{r}_{1}-\mathbf{r}_{3}$ and $\mathbf{r}_{2}-\mathbf{r}_{3}$
(thus eliminating the need to include the former), the vectors $\mathbf{r}_{1}+\mathbf{r}_{2}$,
$\mathbf{r}_{1}+\mathbf{r}_{3}$ and $\mathbf{r}_{2}+\mathbf{r}_{3}$
are linearly independent (the plane made by any two of them will not
contain the third), and thus should all be included.

Finally, the 3pcf in an off-center LTB universe is 
\begin{equation}
\varphi_{w}=\varphi_{w}\left(\left|\mathbf{r}_{1}-\mathbf{r}_{3}\right|,\left|\mathbf{r}_{2}-\mathbf{r}_{3}\right|,\left|\mathbf{r}_{1}+\mathbf{r}_{3}-2\mathbf{w}\right|,\left|\mathbf{r}_{2}+\mathbf{r}_{3}-2\mathbf{w}\right|,\left|\mathbf{r}_{1}+\mathbf{r}_{2}-2\mathbf{w}\right|\right)\,.\label{eq:3pcf-offcenter-ltb}
\end{equation}
This can be obtained from (\ref{eq:3pcf-ltb}) as follows: since the
location of the special point $\mathbf{w}$ is arbitrary, the 3pcf
should satisfy a shift symmetry analogous to (\ref{eq:shift-symmetry}).
We thus shift all points in (\ref{eq:3pcf-ltb}) by an arbitrary amount
$\mathbf{a}$ and $\mathbf{w}$ so as to make the result shift invariant.
This gives the above result.

\section{Final remarks\label{sec:Final-remarks}}

Correlation functions belong to the core of modern cosmology. The
perspective of extending the $\Lambda$CDM model to inhomogeneous,
anisotropic, and non-Gaussian universes depends crucially on our abilities
to model and measure such functions with increasing levels of sophistication.
In this work we have introduced a novel formalism which allows us
to fix the functional dependence of correlation functions given the
underlying spacetime (continuous) symmetries. Given a set of Killing
vectors, we have found a set of first order partial differential equation
which can be solved for the functional dependence of the correlation
function. The method works for arbitrary $N$-point correlators as
long as one stays in the Born approximation – that is, as long as
cosmological perturbations can be treated as external fields in a
fixed background. We have also provided a general solution to the
two-point correlation function which naturally introduces the time
dependence, provided one finds a set of triad vectors commuting the
Killing vector fields. This solution is particularly useful in applications
to Bianchi cosmologies, where such triad of vectors can always be
found \cite{stephani2009exact,Pontzen:2010eg}.

We have successfully applied the formalism to the two-point function
in three different cosmological spacetimes, namely, the anisotropic
and spatially flat solutions of Bianchi type I and VII$_{0}$, and
to the case of an off-center LTB universe, which includes the standard
LTB model as a special case. Specializing to the case of CMB temperature
fluctuations, we have provided asymptotic expansions of these correlation
functions around the known Friedmannian case. Each spacetime leaves
its own multipolar fingerprint on the CMB covariance matrix $\left\langle a_{\ell_{1}m_{1}}a_{\ell_{2}m_{2}}\right\rangle $.
To the lowest order in the expansion parameters, we have found that
Bianchi-I spacetimes lead to quadrupolar couplings $\left\langle a_{\ell_{1}m_{1}}a_{\ell_{1}\pm2,m_{2}}\right\rangle $
while preserving the isotropic angular spectrum $C_{\ell}$. Bianchi
VII$_{0}$ models, on the other hand, lead to quadrupole couplings
as well as suppression of the $C_{\ell}$s, whereas an off-center
LTB metric leads to dipolar couplings and a modification of the $C_{\ell}$s
– the latter depending on a free function which has to be fixed by
solving the photon transport equations in this geometry.

We have also applied the method to infer the functional dependence
of (non-Gaussian) three point correlation functions to the (well-known)
case of a FLRW universe, and also to the case of an off-center LTB
universe. As a byproduct we have found a formal link between the three-point
correlation function in an FLRW universe and a Gaussian 2pcf in an
off-center LTB universe. This link results from the fact that a universe
with a strong dependence on the three-point correlation function is
geometrically degenerate to a Gaussian universe with a special point.

We would like to end with some remarks on the limitations and possible
extensions of the formalism. First we stress that, although the method
can be used to quickly give the CMB multipolar couplings in a given
geometry, it cannot be expected to give more information than that.
The case of Bianchi-I is a clear example. While the quadrupolar couplings
we found here are compatible with the result of more in-depth analysis,
the present formalism cannot predict the oscillations in the power
spectrum resulting from linear perturbation theory \cite{Gumrukcuoglu:2007bx,Pitrou:2008gk}
nor the correlation between scalar and tensor modes arising from the
dynamical couplings with the shear \cite{Pereira:2007yy,Gumrukcuoglu:2010yc}.
Second, we have not considered the case of spin functions, which are
of central importance to the physics of polarization and weak-lensing
of the CMB. The case of vector two-point functions in de Sitter spacetimes
have been addressed in \cite{Allen:1985wd} using a different formalism,
where it was found that it also has the same symmetries of the background
space. In the present formalism this conclusion is not immediate since
equation (\ref{eq:2pcf-coord-inv}), when applied to more general
tensor correlators, will introduce new terms coming from the Lie derivative
of the tensor. We postpone such analysis to future publications. Nonetheless,
we emphasize that the method developed here is general, and can be
equally useful in applications to quantum field theory in curved spacetime.

\acknowledgments This work was supported by Conselho Nacional de Desenvolvimento Tecnológico (CNPq) under grant 485577/2013-5. O.H.M thanks   CAPES for financial support. We also thank Cyril Pitrou for insightful remarks on the final version of this work. 

\appendix
\section{Miscellanea}

We gather here some useful formulae and results which were used in
the main text.

\subsection{Power spectrum and Hankel transform}\label{Appendix:hankel-Plm}

In the examples considered in this work, the correlation function
lacks global rotation symmetry, so that it depends on the vector connecting
two points in the following manner
\[
\xi(\mathbf{r}_{2},\mathbf{r}_{1})=\xi\left(\mathbf{r}_{\pm}\right)\,,\qquad\mathbf{r}_{\pm}=\mathbf{r}_{2}\pm\mathbf{r}_{1}\,.
\]
In this case the power spectrum also becomes a direction-dependent
function of the Fourier vector
\[
_{\pm}P(\mathbf{k})=\int{\rm d}^{3}\mathbf{r}_{\pm}e^{-i\mathbf{k}\cdot\mathbf{r}_{\pm}}\xi(\mathbf{r}_{\pm})\,.
\]
To relate the multipolar coefficients of $P_{\pm}$ to those of $\xi$
we first use Rayleigh's expansion
\begin{equation}
e^{-i\mathbf{k}\cdot\mathbf{r}_{\pm}}=4\pi\sum_{\ell,m}i^{-\ell}j_{\ell}(kr_{\pm})Y_{\ell m}(\hat{\mathbf{k}})Y_{\ell m}^{*}(\hat{\mathbf{n}}_{\pm})\,,\qquad\mathbf{r}_{\pm}=r_{\pm}\hat{\mathbf{n}}_{\pm}\,.\label{eq:Rayleigh}
\end{equation}
Next we decompose both $P_{\pm}$ and $\xi$ into spherical harmonics
and use their orthogonality relation to express the multipolar coefficients
of each function. The result is the Hankel transform of the power
spectrum (see ref. \cite{Abramo:2010za} for its use in cosmology)
\begin{equation}
_{\pm}P_{\ell m}=4\pi i^{-\ell}\int_{0}^{\infty}r_{\pm}{\rm d}r_{\pm}\,j_{\ell}(kr_{\pm})\xi_{\ell m}(r_{\pm})\,.\label{eq:hankel-transform}
\end{equation}

\subsection{Covariance matrix}

Since expression (\ref{eq:cov-matrix-real-space}) is not very popular,
we show here that it does lead to the correct results when the universe
is homogeneous and isotropic. For a FLRW universe we have
\[
\xi_{\ell_{3}m_{3}}(r_{-})=\xi_{00}(r_{-})\delta_{\ell_{3}0}\delta_{m_{3}0}\,.
\]
For this multipolar combination the Gaunt factor becomes
\[
{\cal G}_{m_{1}m_{2}0}^{\ell_{1}\ell_{2}0}=\frac{(-1)^{m_{1}}}{\sqrt{4\pi}}\delta_{\ell_{1}\ell_{2}}\delta_{m_{1},-m_{2}}\,.
\]
Moreover
\begin{align*}
J_{\ell_{1}\ell_{1}0}^{(-)}(r_{-}) & =\int_{0}^{\infty}k^{2}{\rm d}k\,j_{\ell_{1}}(k\Delta\eta)j_{\ell_{1}}(k\Delta\eta)j_{0}(kr_{-})\\
 & =\frac{\pi}{2\left(\Delta\eta\right)^{2}r_{-}}\int_{0}^{\infty}{\rm d}xJ_{\ell_{1}+1/2}^{2}(x)\sin(2ax)\,,\qquad a\equiv r_{-}/\left(2\Delta\eta\right)\\
 & =\frac{\pi}{4\left(\Delta\eta\right)^{2}r_{-}}P_{\ell_{1}}\left(1-2a^{2}\right)
\end{align*}
where in the last step we have used integral 6.672.5 of ref. \cite{jeffrey2007table}.
Next we recall that 
\[
r_{-}^{2}=2\left(\Delta\eta\right)^{2}\left(1-\cos\gamma\right)=4\left(\Delta\eta\right)^{2}a^{2}
\]
which gives
\[
J_{\ell_{1}\ell_{1}0}^{(-)}(r_{-})=\frac{\pi}{4\left(\Delta\eta\right)^{2}r_{-}}P_{\ell_{1}}\left(\cos\gamma\right)\,.
\]
Bringing everything together in expression (\ref{eq:cov-matrix-real-space})
we find
\begin{align*}
\left\langle a_{\ell_{1}m_{1}}a_{\ell_{2}m_{2}}\right\rangle _{-} & =\frac{8}{9}\int_{0}^{2\Delta\eta}r_{-}^{2}{\rm d}r_{-}\,\xi_{00}^{*}(r_{-})J_{\ell_{1}\ell_{1}0}^{(-)}(r_{-})\frac{\left(-1\right)^{m_{1}}}{\sqrt{4\pi}}\delta_{\ell_{1}\ell_{2}}\delta_{m_{1},-m_{2}}\,,\\
 & =\frac{2\pi}{9\left(\Delta\eta\right)^{2}}\frac{(-1)^{m_{1}}}{\sqrt{4\pi}}\left[\int_{0}^{2\Delta\eta}r_{-}{\rm d}r_{-}\,\xi_{00}^{*}(r_{-})P_{\ell_{1}}(\cos\gamma)\right]\delta_{\ell_{1}\ell_{2}}\delta_{m_{1},-m_{2}}\,,
\end{align*}
We now note that $r_{-}{\rm d}r_{-}=\left(\Delta\eta\right)^{2}{\rm d}\left(-\cos\gamma\right)$
so that
\begin{align*}
\left\langle a_{\ell_{1}m_{1}}a_{\ell_{2}m_{2}}\right\rangle  & =(-1)^{m_{1}}\left[\frac{2\pi}{9}\int_{-1}^{1}{\rm d}\left(\cos\gamma\right)\xi_{FL}(\gamma)P_{\ell_{1}}(\cos\gamma)\right]\delta_{\ell_{1}\ell_{2}}\delta_{m_{1},-m_{2}}\,,\\
 & =(-1)^{m_{1}}C_{\ell_{1}}\delta_{\ell_{1}\ell_{2}}\delta_{m_{1},-m_{2}}\,,
\end{align*}
where in the last line we have used eq. (\ref{eq:SW-2pcf}) and $\xi_{FL}(\gamma)=\xi_{00}/\sqrt{4\pi}$.

\subsection{Gaunt coefficients}\label{subsec:gaunt}

The Gaunt coefficients result from the integral of three spherical
harmonics over the sphere. They are given by \cite{edmonds}
\begin{align*}
{\cal G}_{m_{1}m_{2}m_{3}}^{\ell_{1}\ell_{2}\ell_{3}} & =\int{\rm d}^{2}\hat{\mathbf{n}}Y_{\ell_{1}m_{1}}(\hat{\mathbf{n}})Y_{\ell_{2}m_{2}}(\hat{\mathbf{n}})Y_{\ell_{3}m_{3}}(\hat{\mathbf{n}})\\
 & =\sqrt{\frac{\left(2\ell_{1}+1\right)\left(2\ell_{2}+1\right)\left(2\ell_{3}+1\right)}{4\pi}}\left(\begin{array}{ccc}
\ell_{1} & \ell_{2} & \ell_{3}\\
0 & 0 & 0
\end{array}\right)\left(\begin{array}{ccc}
\ell_{1} & \ell_{2} & \ell_{3}\\
m_{1} & m_{2} & m_{3}
\end{array}\right)\,.
\end{align*}
where the $3\times2$ matrices are the Wigner 3-J symbols. The Gaunt
coefficients are identically zero whenever the sum $\ell_{1}+\ell_{2}+\ell_{3}$
is an odd number, and whenever $m_{1}+m_{2}+m_{3}\neq0$.

\subsection{Bipolar power spectrum}\label{subsec:biposh}

The bipolar power spectrum \cite{Hajian:2003qq,Hajian:2005jh} are
the harmonic coefficients of the correlation function when expanded
in a basis of bipolar spherical harmonics \cite{varshalovich1988quantum}.
They are related to the covariance matrix as
\begin{equation}
{\cal A}_{\ell_{1}\ell_{2}}^{LM}=\sum_{m_{1},m_{2}}\left\langle a_{\ell_{1}m_{1}}a_{\ell_{2}m_{2}}\right\rangle (-1)^{M+\ell_{1}-\ell_{2}}\sqrt{2L+1}\left(\begin{array}{ccc}
\ell_{1} & \ell_{2} & L\\
m_{1} & m_{2} & -M
\end{array}\right)\,.\label{eq:biposh-1}
\end{equation}
Using the identity \cite{edmonds}
\[
\sum_{L,M}(2L+1)\left(\begin{array}{ccc}
\ell_{1} & \ell_{2} & L\\
m_{1} & m_{2} & M
\end{array}\right)\left(\begin{array}{ccc}
\ell_{1} & \ell_{2} & L\\
m'_{1} & m'_{2} & M
\end{array}\right)=\delta_{m_{1}m'_{1}}\delta_{m_{2}m'_{2}}\,,
\]
the inverse relation is found to be
\begin{equation}
\left\langle a_{\ell_{1}m_{1}}a_{\ell_{2}m_{2}}^{*}\right\rangle =(-1)^{m_{1}+\ell_{1}-\ell_{2}}\sum_{L,M}\sqrt{2L+1}{\cal A}_{\ell_{1}\ell_{2}}^{LM}\left(\begin{array}{ccc}
\ell_{1} & \ell_{2} & L\\
m_{1} & -m_{2} & -M
\end{array}\right)\,.\label{eq:cov-from-biposh}
\end{equation}
By inserting (\ref{eq:cov-matrix-real-space}) into (\ref{eq:biposh-1})
and using \cite{edmonds}
\[
\sum_{m_{1},m_{2}}\left(\begin{array}{ccc}
\ell_{1} & \ell_{2} & \ell_{3}\\
m_{1} & m_{2} & m_{3}
\end{array}\right)\left(\begin{array}{ccc}
\ell_{1} & \ell_{2} & L\\
m_{1} & m_{2} & -M
\end{array}\right)=\frac{\delta_{L\ell_{3}}\delta_{M,-m_{3}}}{\sqrt{2L+1}}
\]
one arrives at (\ref{eq:biposh}), where the coefficients ${\cal F}_{\ell_{1}\ell_{2}L}$
were defined by
\begin{equation}
{\cal F}_{\ell_{1}\ell_{2}L}=(-1)^{\ell_{1}-\ell_{2}}\sqrt{\frac{\left(2\ell_{1}+1\right)\left(2\ell_{2}+1\right)\left(2L+1\right)}{4\pi}}\left(\begin{array}{ccc}
\ell_{1} & \ell_{2} & L\\
0 & 0 & 0
\end{array}\right)\,.\label{eq:Flll}
\end{equation}

\bibliographystyle{JHEP}
\bibliography{2pcf-references}

\end{document}